# Point-wise posteriori phase estimation in high-precision fringe projection profilometry

**LIU Cong[1*], ZHANG Chuang[1], YIN Zhuoyi[2], LIU Xiaopeng[3], & XU Zhihong[1]**

[1] School of Science, Nanjing University of Science and Technology, Nanjing, Jiangsu 210094, China
[2] School of Civil Engineering, Southeast University, Nanjing, Jiangsu 210094, China
[3] College of Computer Science and Engineering, Shandong University of Science and Technology,
    Huangdao, Shandong266590, China
* E-mail: LiuC@njust.edu.cn

**Abstract**

In fringe projection profilometry, the high-order harmonics information of non-sinusoidal fringes will lead to errors in the phase estimation. In order to solve this problem, a point-wise posterior phase estimation (PWPPE) method based on deep learning technique is proposed in this paper. The complex nonlinear mapping relationship between the multiple gray values and the sine / cosine value of the phase is constructed by using the feedforward neural network model. After the model training, it can estimate the phase values of each pixel location, and the accuracy is higher than the point-wise least-square (PWLS) method. To further verify the effectiveness of this method, a face mask is measured, the traditional PWLS method and the proposed PWPPE method are employed, respectively. The comparison results show that the traditional method is with periodic phase errors, while the proposed PWPPE method can effectively eliminate such phase errors caused by non-sinusoidal fringes.

### 1. Introduction

Fringe projection profilometry (FPP) is widely used for three-dimensional (3D) shape or deformation measurements because of the advantages of high accuracy, large field of view (FOV), and fast measurement speed [1-3].

Generally speaking, in order to obtain a high-precision measurement result, the FPP technique requires that the captured fringe images is with approximately ideal sinusoidal gray distribution. However, many factors may affect the fringe quality, such as gamma distortion, image noise and so on. This makes it difficult to calculate the high-precision phase value without nonlinear calibration and correction [4], the gamma calibration method [5] is proposed to ensure that the captured fringes are with sinusoidal distribution, but the calibration process is a little complicated. Although many nonlinear calibration methods [6-9] have been developed and successfully applied, the problem seems to be much more complicated because the projector's nonlinear gamma effect actually changes over time. A more versatile technique is the digital binary defocusing method [10, 11]. However, it has some limitations that the accuracy depends on whether the defocusing degree is appropriate. To avoid the phase errors caused by improper defocusing, the projected images are optimized, such as random dithering, ordered dithering, error diffusing dithering [12, 13] and the genetic algorithm [14]. However, these methods require a relatively larger fringe period. Regardless of the method adopted above, the problem under consideration is transformed to a complex non-linear solution problem.

At present, artificial intelligence (AI) has made comprehensive breakthroughs in computer vision, image or speech processing, and other fields. At the same time, deep learning technology has also been successfully applied in optical imaging, computational imaging, holographic microscopy, and so on [15-21], it shows great potential and broad application prospects. With multi-layer neural analysis, it can estimate highly nonlinear problem. As long as there are enough neurons in hidden layers of the neural network, the nonlinear dimensional transformation of input and output can be established with simple linear relationships [22]. Therefore, it has been combined to the FPP technology. It is currently applied mostly to Fourier phase extraction methods [23]. However, the precision of the Fourier method is lower than that of the phase-shifting method. There is also an establishment of the relationship between the phase and the height of the object [24, 25]. Cuevas et al. proposed a radial basis function (RBF) neural network model to predict the actual height of objects without knowing the experimental optical parameters [26]. The neural network method is also applied to phase unwrapping [27]. Spoorthi G.E. et al. proposed a phase neural network (PhaseNet) method to achieve two-dimensional spatial phase unwrapping, which has a good noise suppression effect [28]. In addition, it can also be used for phase error compensation. Yang et al. proposed a general phase error compensation method with a three-to-three deep learning framework (Tree-Net), this network can construct the extra three shifted fringe images with the captured three original fringe images [29].

Although the above methods discussed a variety of neural networks in phase calculation and compensation, the highly nonlinear relationship between shifted pixel and phase values seems to remain unsolved. Therefore, this paper proposes a point-wise posterior phase estimation (PWPPE) method based on deep learning. This method solves the issue of the phase errors caused by non-sinusoidal fringe images by establishing an intelligent nonlinear relationship between the pixel gray value and the phase value. It utilizes a flat plane as the calibration object, and achieves accurate phase information in actual experiment through neural network training.

## 2. Principle

The flowchart of PWPPE method proposed in this paper is shown in Figure 1, which mainly includes the following three steps: 1) data acquisition and processing, 2) neural network training, and 3) neural network testing. In the first part, in order to better reflect the real situation, the collected data need to contain several situations from defocusing to focusing. The second part and the third part are supposed to select non-duplicate data from the dataset for training and testing. The principles are discussed in detail as follows:

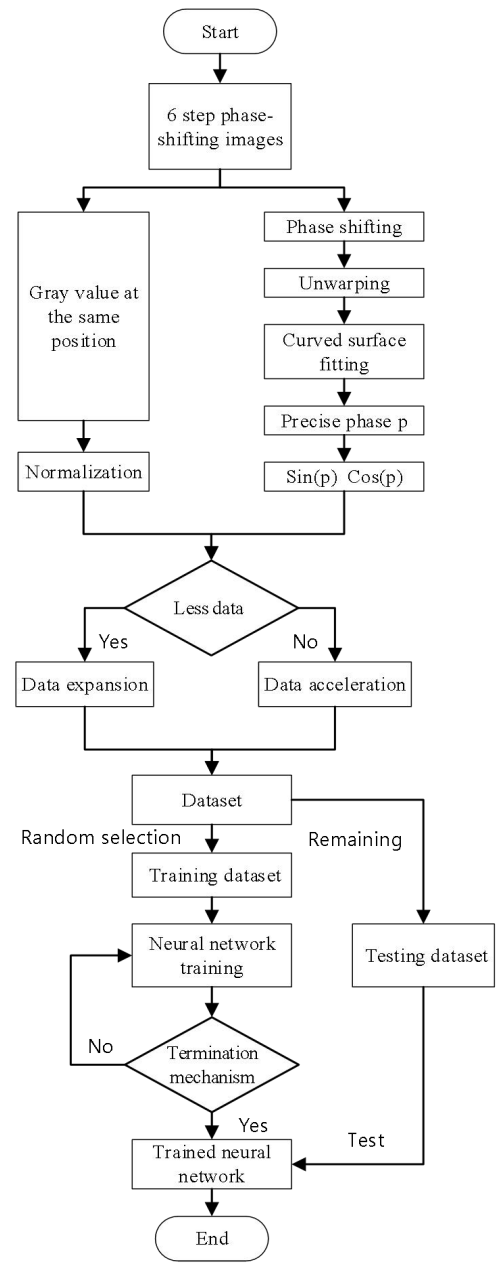

Figure 1 Flowchart of the algorithm

### 2.1. Data acquisition and processing

The formation of dataset includes the following three steps: 1) original data acquisition, 2) data normalization, and 3) data augmentation. The followings describe the formation of the dataset in detail by introducing the input and output process of the neural network:

### 2.1.1. Neural network input

Many factors may introduce the high-order harmonics information to the captured sinusoidal fringe images. Digital binary images defocusing method is commonly employed because binary images can increase the digital light processing (DLP) projection speed, reduce the fringe period, simultaneously expand the FOV of projection. Therefore, the effectiveness of the proposed PWPPE method is verified under different defocusing degrees. However, it is worth mentioning that the application of the PWPPE method is not only limited to this scenario.

In order to simulate the real situation, the real experiment images are applied. Figure 2 shows a captured fringe images in experiments. It can be seen that there is a seam between adjacent fringes, this is because the resolution of the camera is much larger than that of the projector in large FOV measurements. It is hard to be obtained using computer defocusing simulations.

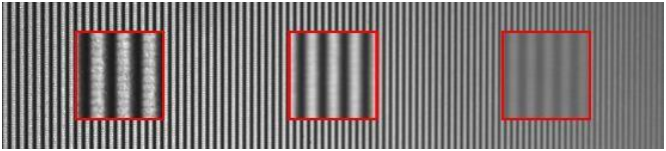

Figure 2 A fringe pattern at different defocusing levels

The captured images should include several situations from defocusing to focusing, as shown in Figure 2. Therefore, a plane calibration plate is set up in the measurement system to project a set of phase-shifted binary fringes. The defocusing degree distribution in the image is significantly different through adjusting the angle of the plate. Limited by the depth of field of the camera, the left side of the images is in focus and the right one is out of focus.

The gray value of a same pixel under different phase shifts are selected to form dataset $D$, $i$ represents the phase shifted step, $D=[D(1),…,D(i),…,D(N)]$. The data vector $D$ is normalized to [-1,1], which can be described as:

$$D_n(i) = \frac{2D(i) - \max(D) - \min(D)}{\max(D) - \min(D)} \quad (1)$$

where $D_n$ is the normalized result of $D$, max and min are the maximum and minimum value of the dataset.

After normalization, the next step is data augmentation. Two optional technical details may be considered.

● If the quantity of data is small, the following data expansion techniques can be used:

$$D_{in}\begin{cases}[D(i+1),D(i+2),\cdots,D(N),D(1),D(2),\cdots,D(i)]\\ [D(i),D(i-1),\cdots,D(1),D(N),\cdots,D(2+i),D(1+i)]\end{cases} \quad (2)$$

where $D_{in}$ represents the new dataset generated by expansion, $i=1,2\cdots N$. Taking the six-step phase-shifting method $N=6$ as an example, one piece of data can be expanded to 12 pieces of data by means of similar rotation and symmetry.

● If the quantity of data is large, the following data acceleration techniques can be adopted.

The largest amount of data in $D$ is selected. Assuming that its location is $j$, it can be expressed as:

$$D_{ac} = [D(j), D(j+1), \cdots, D(N), D(1), D(2), \cdots, D(j-1)] \quad (3)$$

Where $D_{ac}$ represents an improved dataset for accelerating the network training. This output value is limited to $\left[\frac{-2\pi}{N}, \frac{2\pi}{N}\right]$. Note that because of existence of random noise, the selection of maximum value may be difficult; thus, the range of output value is not precise. Once $j$ is recorded, its true phase value can be calculated.

Experiments show that data augmentation has a positive effect on the stability and accuracy of neural network training.

### 2.1.2. Neural network output

Ideally, under the phase-shifting method, the gray value of $i$'th step $I_i$ at each point can be expressed as:

$$A(x,y) + B(x,y)\cos(\phi(x,y) + \theta_i) - I_i(x,y) = 0 \quad (4)$$

Where $(x, y)$ represent the pixel position, $A$ is the background light intensity, $B$ is the surface reflectivit, $\phi$ the corresponding phase value, and $\theta$ the shifted phase. $A$, $B$ and $\phi$ are unknown quantities that need to be solved. A measurement system may contain high-order harmonics. The high-order harmonics in each point of an image are not always the same. At least one parameter $E$ is needed to describe the form of higher-order harmonics. Thus, the function of the pixel point (x, y) may be expressed as:

$$F_{(x,y)}(A, B, \phi, E) = 0 \quad (5)$$

Therefore, the function relationship $F_{(x, y)}$ contain as least four parameters. In other words, to decouple $\phi$ in this equation, at least four data points are needed. Therefore, the step of phase shift is as least four. To suppress the noise, a six-step phase-shifting method is used. The shifted phase of each fringe pattern is $\frac{2\pi}{6}$.

The point-wise least squares (PWLS) phase solution formula for a six-step phase-shifting can be expressed as:

$$\phi(x,y) = -\arctan\left[\frac{\sum_{i=1}^{6} I_i(x,y)\sin\frac{2\pi i}{6}}{\sum_{i=1}^{6} I_i(x,y)\cos\frac{2\pi i}{6}}\right] \quad (6)$$

Where $I_i$ is the gray value corresponding to each step in the phase-shifting method, and $(x, y)$ is the pixel coordinate. The phase value $\phi$ obtained by PWLS method contains errors caused by higher-order harmonics.

After the phase value of each pixel in the image is solved, the unwrapping operation is carried out. Since the image of the collected dataset is a plane calibration plate, the phase distribution after unwrapping should be approximately in accord with the linear function. Because the phase error is regular high frequency [30]. The fitting of image space can effectively eliminate it. So the precise phase value of each pixel position on [-π, π] can be obtained. If there is no plane calibration plate with high machining accuracy, it can also be replaced by any object with stable surface height change.

$$\phi_{in} = \begin{cases} \phi + i\frac{2\pi}{N} & \rightarrow [-\pi, \pi] \\ 2\pi - (\phi + i\frac{2\pi}{N}) & \rightarrow [-\pi, \pi] \end{cases} \quad (7)$$

Where $\phi_{in}$ represents compensated output, and $\rightarrow$ is the value of the phase angle converted to [-π, π].

### 2.2. Neural network training and testing

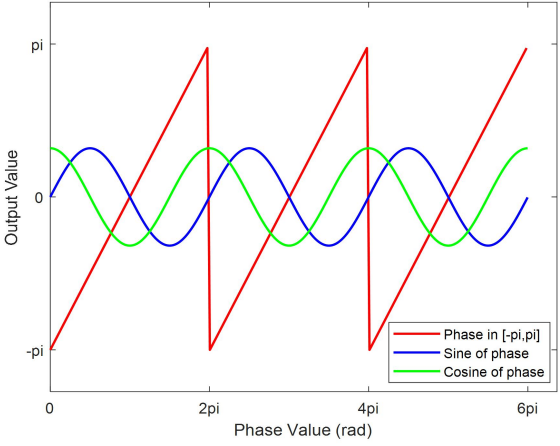

Figure 3 Comparison of the continuity of different output modes

However, the obtained accurate phase value cannot be directly used as the output value. In the previous work, it is found that if only the phase value is taken as a single output value, a sharp boundary effect will appear near -π and π. That is to say, there are serious training convergence errors at -π and π, and the influence of random noise value on the solution results are particularly severe. This phenomenon is due to the fact that if the phase value is directly taken as the output, there is a distance of 2π in the mathematical sense for two values -π and π which are equal in physical sense. That is, there is a mutation that should not exist. Therefore, the output value is replaced by the sine and cosine values corresponding to the phase value. As shown in Figure 3, the sine and cosine values are continuously varying with no abrupt change. This operation is crucial. In later experiments, it is clear that the boundary effect disappears completely. Note that when the sine and cosine values of the phase are known, the phase value can be obtained by an arc tangent operation. This change also has some very wonderful uses. The output values $O_s$ and $O_c$ of the network can be tested to determine whether the solution is correct. The ideal value of $\sqrt{O_s^2 + O_c^2}$ should be 1. When the difference between $\sqrt{O_s^2 + O_c^2}$ and 1 increases, there may be some problems in the input data or the trained network. The established solution system forms a self-test.

It can be seen from the previous section that the input data has been expanded, and the corresponding output set only must compensate for several $\frac{2\pi}{N}$, which are expressed as follows:

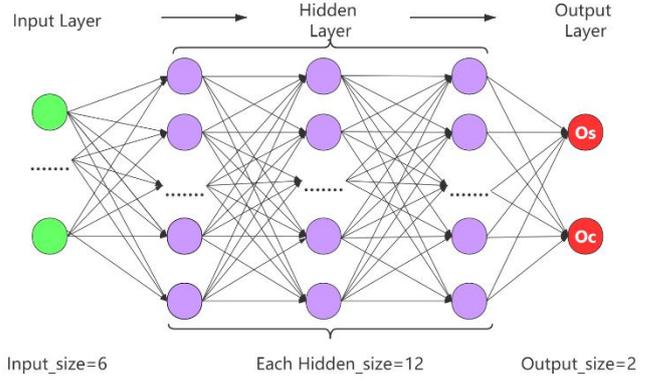

Figure 4 Neural Network Structure Diagram

The following neural network model is developed. And feed-forward neural network (FNN) model based on deep learning is employed. As shown in Figure 4, in addition to the input and output layers, there are three hidden layers and each containing twelve neurons. As shown in Figure 5, in order to increase the nonlinear mapping of the system, the activation function $F(x)$ of each layer of neurons can be expressed as follows:

$$F(x) = \frac{2}{1 + e^{-2x}} - 1 \quad (8)$$

where *x* is the value received by each neuron.

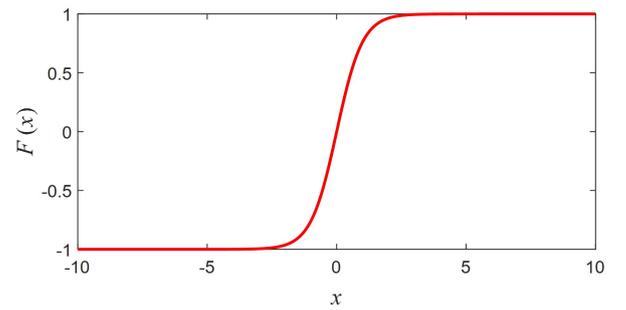

Figure 5 Activation function

In this paper, the dataset is collected by the method mentioned above and randomly scramble them. One part of the dataset is selected as the training dataset, and the other part as the test dataset. GPU (NVIDIA Quadro P3200) and CPU (i7-8850H) muti-threading are used to accelerate the neural network training. After 10,000 iterations, the network converged to the minimum error, and the mean-square error is less than $5 \times 10^{-4}$ rad. In the additional test, only 1% random points are sampled in the test dataset, similar results can be achieved. Thus, the random sampling method can extremely reduce training costs. Only 10 minutes were required to complete 10,000 training iterations.

## 3. Simulations

In this paper, the plane calibration plate is used as the simulation object. Firstly, the calibration plate, the camera and the projector are fixed on the optical platform to limit the relative displacement between the parts. In this process, the camera lens direction should be pointed to the plane of the calibration plate, and the calibration plate should be in the middle position in the camera perspective. The projector lens should be aligned to the maximum plane direction of the calibration plate for positioning. Then, the analog dataset is collected and processed. The processed dataset is input into the trained neural network. The phase error is shown in Figure 6. The errors of PWPPE method are basically within $\pm 0.03$ rad, while the errors of PWLS method basically reach 0.1 rad in the first 500 pixels, and the errors from 500 pixels onwards are basically the same as the method in this paper. This shows that the PWPPE method is not only accurate but also more stable.

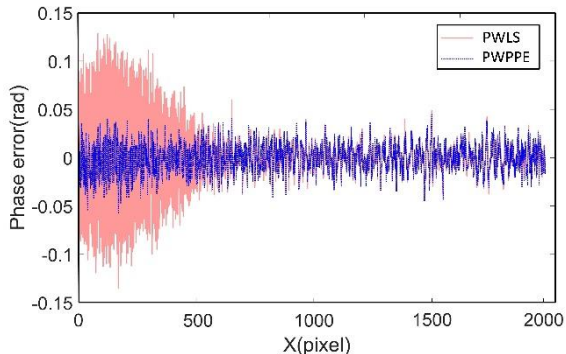

Figure 6 Result of one row in trained image

The generalization ability of the neural network is examined next. While keeping the relative positions of projector and camera unchanged, the position of the calibration plate and the focal length of the camera or projector are changed. The first group keeps the position of the calibration plate unchanged and changes the focal length. In groups 2 and 3, the focal length is kept constant and the position of the planar calibration plate in space is adjusted. In group 4, the other settings remain unchanged, reducing the exposure time or aperture.

Table 1. Mean-square error of phasemeasurement( $10^{-4}$ ) rad

|      | Trained | Group 1 | Group 2 | Group 3 | Group 4 |
|------|---------|---------|---------|---------|---------|
| PWLS | 4.626 | 3.912 | 5.168 | 5.233 | 8.974 |
| PWPPE | 2.3374 | 2.4541 | 2.319 | 2.961 | 3.086 |

PWLS and PWPPE methods are used to calculate the phase of the plane calibration plate. As shown in Table 1, the mean-square errors of the two methods in four groups are obtained. It can be clearly seen that the mean-square error of PWPPE method for solving the four groups of results are smaller, and the stability of the mean-square errors are higher. Therefore, compared with the PWLS method, the PWPPE method not only has a great improvement in accuracy, but also has a relatively stable solution accuracy. At the same time, in order to ensure that the measurement accuracy reaches the limit, the saturation of the light intensity in the image must be ensured. When the spatial position changes, the accuracy will decrease slightly. In the actual measurement, the calibration plane data of multiple positions can be obtained simultaneously for training, which greatly improves the spatial robustness of phase calculation.

Compared with the phase correction method combined with local image, the proposed method only needs the phase shift data of a single pixel to obtain the phase value. This means that this method has higher independent resolution. For the measured object with large gradient or the image with large variation of high-order harmonics coefficient of non-sinusoidal wave, the method combined with local image may face the risk of reducing the accuracy or even failure. Taking Pan Bing's method [30] as an example, because it needs to solve an iterative parameter c with the help of the whole image, when the high-order harmonics coefficient of non-sinusoidal wave in the image changes greatly, the parameter is not a constant for the whole measurement image. The image in Figure 2 is taken as an example, and the solution results are shown in Figure 7. This means that the risk of parameter deviation is faced in the process of correction.

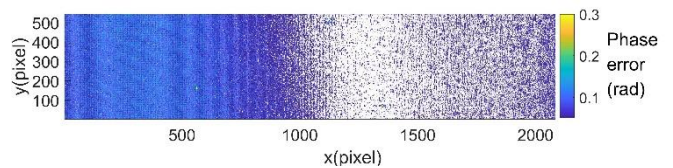

Figure 7 Calculation results of method [30].

## 4. Experiments

To further verify the effectiveness of the proposed method,

a face mask is used as the experimental object in Figure 8. The entire experimental system uses a DLP6500FLQ projector and a UI-3370CP-M-GLRev-2 camera. The resolution of the camera is 2048 × 2048 pixels and frame rates of 80 fps. The camera is configured with a 35 mm focal length Kowa lens.

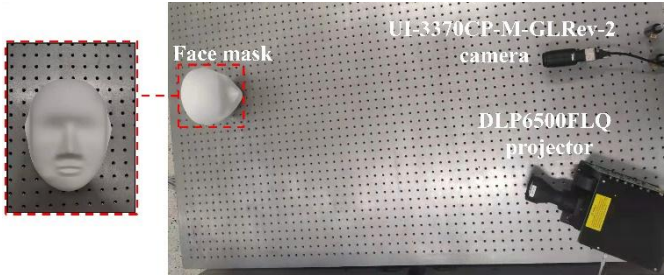

Figure 8 Experimental apparatus.

The projector projected 6 fringe images into the face mask. The camera is focused directly on a point on the surface of the object. The image corresponding to the six-step phase-shifting method is obtained. A group of phase-shifting images can generate up to 2048×2048×12 different pieces of data, which is considerable.

## 5. Results and discussion

PWLS and PWPPE method to solve the phase of face mask respectively. Figure 9 (a), (b) and (c) respectively represent the real image of the face mask, the image obtained by PWLS method and the image obtained by PWPPE method. Figure 9 (d) and (e) show the phase obtained by the two methods at the position of the red mark line. The phase height near the pixel point 1000 is selected and magnified, and it is obvious that the phase height curve obtained by the PWLS method fluctuated significantly, indicating that the results obtained by this method has periodic phase errors. However, the phase height curve obtained by PWPPE method is smooth, which indicates that this method can well suppress such errors and has higher precision. At the same time, the validity of the method can also be judged by the self-test technique. As shown in Table 2, 74% of the results have errors of $\pm 0.01$, 94% of the results with errors of $\pm 0.1$, and the maximum phase errors are $\pm 0.12$. The results obtained by this method are basically consistent with the ideal results, and the fluctuation range is within the allowable range of random noise. It can be considered that this method can effectively solve the above problems, and its accuracy is higher than that of PWLS method.

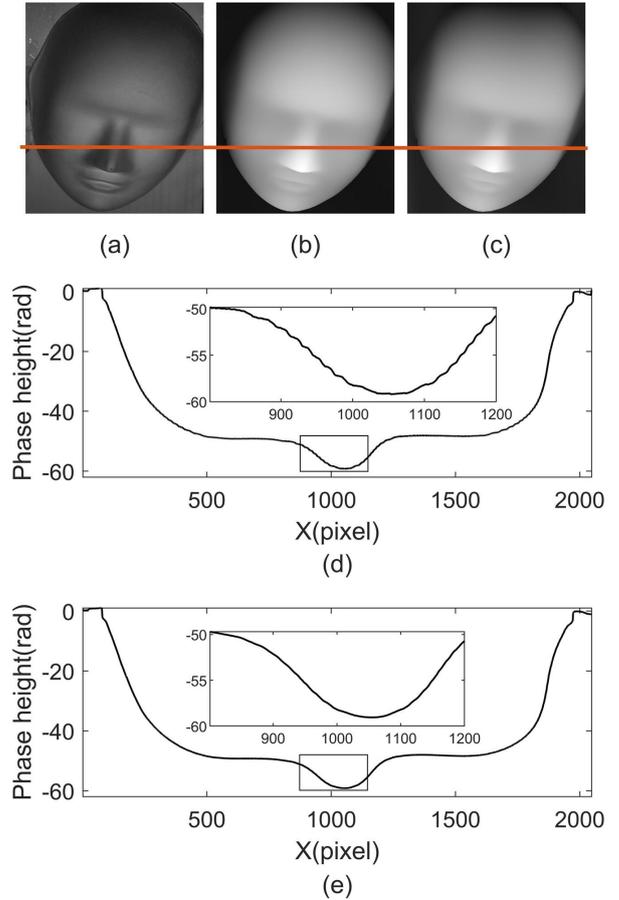

Figure 9 Experimental results of face measurement. (a) Object image; (b) result of PWLS method; (c) result of PWPPE method; (d) result of one row with PWLS method and filtering; (e) result of one row with PWPPE method.

**Table 2. Self verification**

| $\sqrt{O_s^2 + O_c^2}$ | Proportion |
|---|---|
| $1 \pm 0.12$ | $\approx 100\%$ |
| $1 \pm 0.1$ | 96% |
| $1 \pm 0.05$ | 86% |
| $1 \pm 0.01$ | 74% |

## 6 Conclusion

The proposed method (PWPPE) can estimate the phase height information of the measured object only by obtaining the gray value of the phase shift image. The proposed method can effectively suppress the phase period errors, and its accuracy is higher than that of traditional methods. This method only needs to be calibrated when the equipment is used for the first time, and there is no need to introduce new steps in the measurement process during subsequent measurements. This method does not affect the frame rate of measurement and complicate the measurement process. It can completely eliminate the uncertainty in measurement accuracy via the defocusing band. In the measurement process, if the method proposed in this paper is used, the camera will focus on the surface of the object to be measured. Then, the exposure time or aperture is adjusted to make the image brightness as saturated as possible. The highest measurement accuracy can be achieved by executing the above two steps. The only drawback of this method is that it takes more time to solve the phase. On one hand, this can be easily overcome by hardware acceleration, such as through use of a GPU. On the other hand, in most cases, the importance of accuracy is far greater than calculation time.

## Acknowledgements

This work was supported by the National Natural Science Foundation of China (Grant No.11802132), the Natural Science Foundation of Jiangsu Province (Grant No. BK20180446), the China Postdoctoral Science Foundation (Grant No. 2020M671493, and 2019M652433), the Jiangsu Planned Projects for Postdoctoral Research Funds, and the Natural Science Foundation of Shandong Province (Grant No. ZR2018BF001).